\documentclass[twoside,a4paper]{article}
\usepackage{qic,epsfig}
\usepackage{graphics}
\usepackage{latexsym}

\newtheorem{thm}{Theorem}
\newtheorem{cor}{Corollary}
\newtheorem{lemm}{Lemma}
\newtheorem{pro}{Proposition}
\newtheorem{de}{Definition\rm}
\newtheorem{ob}{Observation}

\newtheorem{rem}{Remark}

\begin{document}
\setlength{\textheight}{8.0truein}    %FOR 2ND PAGE ONWARDS

\runninghead{Mathematical framework for detection $\ldots$}
            {A. SaiToh, R. Rahimi, M. Nakahara}

\normalsize\textlineskip
\thispagestyle{empty}
\setcounter{page}{1}

%\copyrightheading{Vol.}{No.}{Year}{Page Nos.}
\copyrightheading{0}{0}{2003}{000--000}

\vspace*{0.88truein}

\alphfootnote

\fpage{1}

\centerline{\bf
%%%%%%%%%%%%%%%%%%%%%
%Put in titiles here
%%%%%%%%%%%%%%%%%%%%%
Mathematical framework for detection and quantification}
\vspace*{0.035truein}
\centerline{\bf of nonclassical correlation}
%\vspace*{0.035truein}
%\centerline{\bf FOR QUANTUM INFORMATION AND COMPUTATION\footnote{Typeset the
%title in 10 pt Times Roman, uppercase and boldface.}}
\vspace*{0.37truein}
\centerline{\footnotesize
%%%%%%%%%%%%%%%%%%%%%%%%%%%%%%%%%%%%
%put authors' name and address here
%%%%%%%%%%%%%%%%%%%%%%%%%%%%%%%%%%%%
AKIRA SAITOH$^1$,
~ROBABEH RAHIMI$^{1,2}$,
~MIKIO NAKAHARA$^{1,3}$}
\vspace*{0.015truein}
\centerline{\footnotesize\it
$^1$Research Center for Quantum Computing, Interdisciplinary}
\baselineskip=10pt
\centerline{\footnotesize\it
Graduate School of Science and Engineering, Kinki University}
\baselineskip=10pt
\centerline{\footnotesize\it 3-4-1 Kowakae, Higashi-Osaka, Osaka 577-8502, Japan}
\centerline{\footnotesize\it
$^2$Department of Chemistry and Materials Science}
\baselineskip=10pt
\centerline{\footnotesize\it
Graduate School of Science, Osaka City University}
\baselineskip=10pt
\centerline{\footnotesize\it
3-3-138 Sugimoto, Sumiyoshi-ku, Osaka 558-8585, Japan}
%\vspace*{10pt}
%\vspace*{0.015truein}
\centerline{\footnotesize\it
$^3$Department of Physics, Kinki University}
\baselineskip=10pt
\centerline{\footnotesize\it 3-4-1 Kowakae, Higashi-Osaka, Osaka 577-8502, Japan}
\vspace*{0.225truein}
\publisher{(received date)}{(revised date)}
\vspace*{0.21truein}

%% \abstracts{first paragraph}{second paragraph}{third paragraph}
%% If there is only one paragraph, just keep the second and third empty 
%% like the following one 
\abstracts{
%%For the reviewing process the baselineskip is enlarged.
%%\baselineskip=15pt
%%%%%%%%%%%%%%%%%%%%
% put abstract here
%%%%%%%%%%%%%%%%%%%%
Existing measures of bipartite nonclassical correlation that is typically
characterized by nonvanishing nonlocalizable information under the
zero-way CLOCC protocol are expensive in computational cost.
We define and evaluate economical measures on the basis of a new class
of maps, eigenvalue-preserving-but-not-completely-eigenvalue-preserving
(EnCE) maps. The class is in analogy to the class of
positive-but-not-completely-positive (PnCP) maps that have
been commonly used in the entanglement theories. Linear and nonlinear
EnCE maps are investigated. We also prove subadditivity of the measures
in a form of logarithmic fidelity. 
}{}{}

\vspace*{10pt}

\vspace*{3pt}
\communicate{to be filled by the Editorial}

\vspace*{1pt}\textlineskip    %) USE THIS MEASUREMENT WHEN THERE IS
   %) A SECTION HEADING
%\vspace*{-0.5pt}
%\noindent
%%%%%%%%%%%%%%%%%%%%%%%%%%%%%%%%
%put the text of the paper here
%%%%%%%%%%%%%%%%%%%%%%%%%%%%%%%%
%PACS
%03.65.Ud       Entanglement and quantum nonlocality
%03.67.-a       Quantum information
%05.30.Ch       Quantum ensemble theory
%02.60.-x       Numerical methods (mathematics)
%03.67.Mn       Entanglement measures, witnesses, and other characterizations
%03.65.Ta       Foundations of quantum mechanics; measurement theory

%%For the reviewing process the baselineskip is enlarged.
%%\baselineskip=18pt

%%%
%  Introduction
%%%
\section{Introduction}
Nonclassical correlation of a bipartite system is an essential resource
to perform quantum information processing \cite{G99,NC2000}.
Entanglement, namely the degree of inseparability, of a system is the
most well-known nonclassical correlation. Besides the entanglement
paradigm, several different paradigms \cite{B99,Z02,O02} have been
proposed in which the set of the states with nonclassical correlation
includes certain nonentangled states.
%%%
Unlike entangled states that are defined on the basis of the
impossibility of preparation under local operations and classical
communications (LOCC) \cite{PV06}, nonclassically correlated states in
the different paradigms have been defined on the basis of
post-preparation stages.
%%%

Historically, nonlocality about locally nonmeasurable separable states
was discussed by Bennett {\em et al.} \cite{B99}. Later, in the context
of system-apparatus correlation, a measure called quantum discord was
defined by Ollivier and Zurek \cite{Z02}, which is a discrepancy of two
expressions of a mutual information that are equivalent in the regime of
classical information theory. Recently, the term quantum discord often
indicates the minimized one over the possible local (orthogonal) projection
sets and is widely used as a measure of nonclassical correlation.
As a typical example that justifies the computational power of nonclassical
correlation other than entanglement, it was reported by Datta
{\em et al.} \cite{DSC08} that the Knill and Laflamme's fast estimation of
a normalized trace of a unitary matrix, which uses a single pseudo-pure
qubit and the remaining qubits in a maximally mixed state \cite{KL98},
exhibits a large quantum discord and vanishing entanglement.
Another well-known definition of nonclassical correlation is the
quantum deficit proposed by Oppenheim {\em et al.} \cite{O02}, which is
nonlocalizable information under the closed LOCC
(CLOCC) protocol that allows only local unitary operations and
operations to send subsystems through a complete dephasing channel.
Among different setups they considered \cite{H05}, the zero-way
setting|namely a setting in which the players are allowed to communicate
under CLOCC only after local complete dephasing possibly subsequent to
local unitary operations|was connected to a mathematically simple
classical/nonclassical separation.

The quantum deficit for a density matrix $\rho^{\rm AB}$ of a bipartite
system ${\rm AB}$ is defined as \cite{H05}
\[
{\rm min}_{\Lambda\in{\rm CLOCC}}
[S_{\rm vN}({\rm Tr_B}\Lambda \rho^{\rm AB})
+ S_{\rm vN}({\rm Tr_A}\Lambda\rho^{\rm AB})]
- S_{\rm vN}(\rho^{\rm AB}),
\] where $S_{\rm vN}(\cdot)$ is the
von Neumann entropy. Here, the system may be kept by a single player
after the process $\Lambda$, i.e., ${\rm Tr_B}\Lambda \rho^{\rm AB}$
or ${\rm Tr_A}\Lambda\rho^{\rm AB}$ possibly becomes a null state.
The entropy of a null state is defined to be zero.
In case of the zero-way setting, the minimum is
obtained when a single player possesses the total system after the
process. Thus the zero-way quantum deficit is equal to a minimum
discrepancy between $S_{\rm vN}[
(\mathcal{V}^{\rm A}\otimes \mathcal{V}^{\rm B}) \rho^{\rm AB}]$
and $S_{\rm vN}(\rho^{\rm AB})$
where $\mathcal{V}^{\rm A, B}$ is a dephasing operation (acting on
a local system) deleting the off-diagonal elements of a target matrix
on a certain local basis; the minimum is taken over all local bases.

The zero-way quantum deficit vanishes if the state has a product
eigenbasis. A quantum bipartite system is called (properly)
classically correlated \cite{H05} if and only if it is
described by a density matrix having a product eigenbasis (PE),
\begin{equation}\label{eq_rho_prod}
 \rho^{\rm AB}_{\rm PE}=\sum_{j,k=1,1}^{d^{\rm A},d^{\rm B}}
e_{jk}|v_j^{\rm A}\rangle\langle v_j^{\rm A}|
\otimes |v_k^{\rm B}\rangle\langle v_k^{\rm B}|,
\end{equation}
where $d^{\rm A}$ ($d^{\rm B}$) is the dimension of the Hilbert space of
 A (B), $e_{jk}$ is the eigenvalue of $\rho$ corresponding to the
 eigenvector $|v_j^{\rm A}\rangle\otimes |v_k^{\rm B}\rangle$.
Thus, a quantum bipartite system consisting of subsystems A and B
is nonclassically correlated if and only if it is described by a density
matrix having no product eigenbasis.

Using this simple classical/nonclassical separation, other measures 
\cite{G07,SRN08} were later proposed.
In particular,  Piani {\em et al.} \cite{PHH08} recently designed
a measure which vanishes if and only if a state has a product eigenbasis.
It is in a similar form as the quantum discord \cite{Z02} and defined as a
distance of two different quantum mutual informations that is minimized
over local maps associated with local positive operator-valued
measurements \cite{D78}. It has also been known \cite{Z02} that the
quantum discord vanishes for both of the system-apparatus and
apparatus-system settings if and only if a state has a product eigenbasis.

A pending problem is that the original nonlocalizable information and
the Piani {\em et al.}'s measure both require expensive computational
tasks to take minimums over all possible local operations. A similar
difficulty exists in Groisman {\em et al.}'s measure \cite{G07}
which is a discrepancy between an original state $\rho^{\rm AB}$ and
the state after dephasing under an eigenbasis of ${\rm
Tr_B}\rho^{\rm AB}\otimes {\rm Tr_A}\rho^{\rm AB}$ (called Schmidt
basis). In fact, the obviously classically correlated state
$(|00\rangle\langle00|+|11\rangle\langle 11|)/2$ is
mapped to $I/4$ by dephasing when the improper Schmidt basis
$\{|0\rangle,|1\rangle\}\otimes\{|\pm\rangle\}$
with $|\pm\rangle=(|0\rangle\pm|1\rangle)/\sqrt{2}$ is chosen
while it is mapped to itself when a proper Schmidt basis (the
computational basis in this case) is chosen.
Thus a minimization over possible Schmidt bases is required.
The recently-proposed measurement-induced disturbance measure 
\cite{L08,D09} is a variant of the Groisman {\em et al.}'s measure;
the same problem exists.
In general, the measures involving a minimization over local operations
are intractable in view of computational cost.

In our previous contribution \cite{SRN08}, an entropic measure $G$
based on a sort of game to find the eigenvalues of a reduced density matrix
from the eigenvalues of an original density matrix was proposed, in the
context of $m$-partite $m$-split nonclassical correlation. This measure
can be computed within a finite time although it does not have a perfect
detection range. Its computational cost is indeed less than those for
the intractable measures, but still exponential in the dimension of
the Hilbert space.

Here we introduce a new class of maps to define measures with
improved computational cost, in the context of bipartite splitting.
It is the class of
eigenvalue-preserving-but-not-completely-eigenvalue-preserving (EnCE)
maps. We find it analogous to the class of
positive-but-not-completely-positive (PnCP) maps \cite{P96,H96,H97}
that are popularly used for detection and quantification of
entanglement. The idea of introducing the class EnCE was briefly mentioned 
in our previous contribution \cite{SRN08-2}. Here, we give mathematically
strict definitions and show the fact that any linear EnCE map is a
concatenation of unitary and anti-unitary operations. Thus the
restriction of the theory is clarified. We further introduce a
measure using a nonlinear EnCE map in order for achieving a wider
detection range.

The measures we propose here on the basis of EnCE maps are not as strong
as those using an infinite number of trials, in the detection range of
nonclassically correlated states. We propose a simple way to
relax this drawback: The detection range is improved by introducing an
average of multiple measures whose detection ranges are mutually
different. This approach is described in Section\ \ref{sec4-4}.

This paper is organized as follows. We start with an overview of
the conventional theory of PnCP maps in Section\ \ref{sec2}. We then
define and evaluate new classes of maps in Section\ \ref{sec3}.
The measures are defined and their properties, such as
subadditivity, are verified in Section\ \ref{sec4}. First non-subadditive
measures are introduced in Section\ \ref{sec4-1}. Second subadditive measures
are introduced in Section\ \ref{sec4-2} with the proof of the subadditivity.
The computational complexity of the subadditive measures is investigated
in Section\ \ref{sec4-3}.
A simple way to relax the drawback in the detection range
of the measures is shown in Section\ \ref{sec4-4}.
Section\ \ref{sec6} summarizes our results with some remarks.
%%%%
% Main part
%%%%

\section{Conventional theory of PnCP maps}\label{sec2}
Quantum physics is governed by completely positive (CP) maps.
Any map which is not CP (nCP) is considered to be physically unfeasible.
There is, however, a class of nCP maps which are useful for
characterizing entanglement. These maps are in the class of
positive-but-not-completely-positive (PnCP) maps. 
It has been more than a decade since the Peres-Horodecki criterion
opened the mathematical study of PnCP maps \cite{P96,H96,H97}.
A PnCP map $\Lambda_{\rm PnCP}$ is positive when acting as a global
operator but nonpositive when acting as $I\otimes\Lambda_{\rm PnCP}$
on a system. It maps a separable state
$\rho_{\rm sep}=\sum_i w_i\rho_i^{\rm A}\otimes \rho_i^{\rm B}$
of a bipartite system AB with nonnegative weights $w_i$ to
a certain (physically feasible) state, while it does not necessarily map
an inseparable state to a positive Hermitian matrix. Thus one finds
a density matrix inseparable if one detects a negative eigenvalue of the
matrix obtained after applying $I\otimes\Lambda_{\rm PnCP}$ to the
density matrix. The PnCP map theory has gathered a broad interest in
relation to detecting entanglement (See, {\rm e.g.}, Ref.~\cite{H09review}).

One might be curious to find an analogue of the PnCP map theory to
detect nonclassical correlations often defined in different ways
\cite{B99, Z02, O02} than that of entanglement. We pursue the
analogous theory to detect nonclassical correlation in the context of
classical/nonclassical separation given by the existence/absence of a product
eigenbasis of a bipartite state.

%
%Definitions
%
\section{Introduction of unconventional classes of maps and
their use}\label{sec3}
We aim to introduce an analogy of the PnCP map theory to the present
paradigm of classical/nonclassical separation. For this purpose,
we define our new classes of maps. Let us start with a linear map theory.
\begin{de}
An eigenvalue-preserving (EP) map
$\Lambda_{\rm EP}$ is a map 
acting on a general $d\times d$ density matrix
$\rho=\sum_{k=1}^d e_k|v_k\rangle\langle v_k|$
(here, $e_k$ and $|v_k\rangle\langle v_k|$ are the $k$th eigenvalue and the
corresponding projector, respectively) such that
\[
 \rho=\sum_{k=1}^d e_k|v_k\rangle\langle v_k|
\begin{array}{c}
{}_{\Lambda_{\rm EP}}\\
\mapsto\\~
\end{array}
\rho'=\sum_{k=1}^d e_k|{v'}_k\rangle\langle {v'}_k|,
\]
where $\{|v_k\rangle\}_k$ and $\{|{v'}_k\rangle\}_k$ are both
complete orthonormal systems (CONSs). The dimension of $\rho'$ is equal
to $d$.
\end{de}
Alternatively, we may define the EP map in the following way:
\begin{de}
An EP map $\Lambda_{\rm EP}$ acting on a quantum system S is a bijection
between the set of projectors $\{|v_k\rangle\langle v_k|\}_{k=1}^d$
generated from the vectors $|v_k\rangle$ of a CONS to the set of
projectors $\{|{v'}_k\rangle\langle {v'}_k|\}_{k=1}^d$ generated from
the vectors $|{v'}_k\rangle$ of a CONS
for any CONS $\{|v_k\rangle\}_{k=1}^d$ of the Hilbert space of S.
\end{de}

A class of EP maps analogous to CP is defined as follows. 
\begin{de}
An EP map $\Lambda$ is a complete EP (CEP) map
if and only if $I\otimes\Lambda$ is also an EP map for identity map $I$
of arbitrary dimension. We denote such $\Lambda$ as $\Lambda_{\rm CEP}$.
\end{de}
\begin{ob}{\rm
One of the simplest CEP maps is
$\widetilde{\mathcal{U}}(d):\rho\rightarrow \tilde u\rho \tilde{u}^\dagger$
where $\tilde u$ is an element of a flag manifold
$\widetilde{\rm U}(d)={\rm U}(d)/{\rm U}(1)^{\times d}$
(here, ${\rm U}(d)$ is the $d$-dimensional unitary group).
}\end{ob}
We now define the PnCP analogy in the following way.
\begin{de}\label{defEnCE}
An EP map $\Lambda$ is an EP-but-not-completely-EP (EnCE) map
if and only if there exists an identity map $I$ of some dimension,
for which $I\otimes\Lambda$ is not an EP map.
We denote such $\Lambda$ as $\Lambda_{\rm EnCE}$.
\end{de}

We have defined a class of EnCE maps.
As is analogous to the usage of a PnCP map, the usage of a linear EnCE
map is to find a certain change of eigenvalues of a density matrix by
applying the map to a local subsystem. This is based on the following
proposition.
%%%Propositions and Proofs
\begin{pro}\label{pro2}Both
$I^{\rm A}\otimes\Lambda_{\rm EnCE}^{\rm B}$ and
$\Lambda_{\rm EnCE}^{\rm A}\otimes I^{\rm B}$
preserve the eigenvalues of a density matrix of a system {\rm AB}
if the density matrix has a product eigenbasis.
\end{pro}
\proof{
Let the density matrix with a product eigenbasis
$\{|v_j^{\rm A}\rangle|v_k^{\rm B}\rangle\}_{j,k=
1,1}^{d^{\rm A},d^{\rm B}}$
be
\[
\sigma^{\rm AB}=
\sum_{j,k=1,1}^{d^{\rm A},d^{\rm B}}
e_{jk}|v_j^{\rm A}\rangle\langle v_j^{\rm A}|
   \otimes |v_k^{\rm B}\rangle\langle v_k^{\rm B}|,
\]
where $e_{jk}$ is the $(jk)$th eigenvalue corresponding to the
$(jk)$th projector $|v_j^{\rm A}\rangle\langle v_j^{\rm A}|
   \otimes |v_k^{\rm B}\rangle\langle v_k^{\rm B}|$.
By the definition, any linear EP map acting
as a local operation should map an eigenbasis of the reduced
density matrix of a target subsystem to another CONS.
Therefore, it is obvious that
\[
(I^{\rm A}\otimes\Lambda_{\rm EnCE}^{\rm B})\sigma^{\rm AB}
=\sum_{j,k=1,1}^{d^{\rm A},d^{\rm B}}
e_{jk}|v_j^{\rm A}\rangle\langle v_j^{\rm A}|
\otimes |{v'}_k^{\rm B}\rangle\langle {v'}_k^{\rm B}|,
\]
where $\{|{v'}_k^{\rm B}\rangle\}_k$ is a CONS of the Hilbert space of B,
which may be different from $\{|v_k^{\rm B}\rangle\}_k$.
It is trivial to show the same proof applies to
$\Lambda_{\rm EnCE}^{\rm A}\otimes I^{\rm B}$.
}%%%end of proof
\begin{cor}\label{corX}
A density matrix $\rho^{\rm AB}$ has no product eigenbasis
if either $(I^{\rm A}\otimes\Lambda_{\rm EnCE}^{\rm B})\rho^{\rm AB}$
or $(\Lambda_{\rm EnCE}^{\rm A}\otimes I^{\rm B})\rho^{\rm AB}$
has eigenvalues different from those of $\rho^{\rm AB}$.
\end{cor}
\proof{
This is the contraposition of Proposition\ \ref{pro2}.
}%end of proof

There is, however, a restriction in the type of the linear
EnCE maps according to the following proposition.
This restriction is later relaxed by a
nonlinear EnCE map.
\begin{pro}\label{pro1}
Any linear EP map can be decomposed into unitary transformations
and a transposition. Hence any linear EnCE map can be decomposed
into unitary transformations and a transposition.
\end{pro}
\proof{
Consider a linear EP map $\Lambda_{\rm lin}$ and two pure states
$|x\rangle\langle x|$ and $|y\rangle\langle y|$ ($|x\rangle$ and
$|y\rangle$ can be nonorthogonal to each other).
Let $|x'\rangle\langle x'|=\Lambda_{\rm lin}(|x\rangle\langle x|)$
and $|y'\rangle\langle y'|=\Lambda_{\rm lin}(|y\rangle\langle y|)$.
Consider a state $\tau=|x\rangle\langle x|+|y\rangle\langle y|$
represented under a certain CONS.
As $\Lambda_{\rm lin}$ changes this CONS to a certain CONS,
$\langle x|\tau|x\rangle$ is equal to
$\langle x'|\Lambda_{\rm lin}(\tau)|x'\rangle$.
This suggests that $|\langle x|y\rangle|=|\langle x'|y'\rangle|$ since
$\Lambda_{\rm lin}$ is a linear map.
Note that this is true for any linear EP map $\Lambda_{\rm lin}$ and
any two pure states $|x\rangle$ and $|y\rangle$.
Therefore, by Wigner's unitary-antiunitary theorem \cite{Wig},
there are only two possible types for $\Lambda_{\rm lin}$, namely,
unitary and antiunitary \cite{Wig2} transformations acting on a target
density matrix. Hence the proposition holds.
}%end of proof 

\begin{ob}\label{ob2}{\rm
A unique nontrivial linear EnCE map is the transposition
$\Lambda_{\rm T}$, according to the above proposition.
%It can be used for detecting nonclassical correlation:
%the map $I\otimes \Lambda_{\rm T}$ obviously preserves the
%eigenvalues for a bipartite state in the form of $\rho_{\rm PE}^{\rm AB}$
%while it may not for a state possessing nonclassical correlation.
As an example of detecting a nonclassical correlation, consider the
density matrix of a two-qubit pseudo-entangled (PS) state,
\begin{equation}\label{eqps}
 \rho_{\rm ps}=(1-p)I/4+p|\psi\rangle\langle\psi|
\end{equation}
with $|\psi\rangle=(|00\rangle+|11\rangle)/\sqrt{2}$ and $0< p \le 1$.
It has a nondegenerate eigenvalue 
$(1+3p)/4$ and a degenerate eigenvalue
$(1-p)/4$ with multiplicity 3.
Its partial transposition,
$(I\otimes\Lambda_{\rm T})\rho_{\rm ps}$, has a nondegenerate eigenvalue
$(1-3p)/4$ and a degenerate eigenvalue $(1+p)/4$ with multiplicity 3.
These two sets of eigenvalues are different for $p>0$, indicating
the existence of a nonclassical correlation.

It should be noted that having different eigenvalues after partial
transposition is a sufficient but not necessary condition for a state
to have no product eigenbasis. For example, a 2-qubit state
\begin{equation}\label{eq0+}
\rho_{0+}=\frac{1}{2}(|00\rangle\langle00|+|++\rangle\langle++|)
\end{equation}
with $|+\rangle=(|0\rangle+|1\rangle)/\sqrt{2}$ has no product
eigenbasis because $|0\rangle\langle 0|$ and $|+\rangle\langle +|$
cannot be diagonalized simultaneously.
It is clear that the partial transposition does not change the state and
hence it does not detect a nonclassical correlation.
We may use, instead, the nonlinear map $\mathcal{P}_x$ defined later
in order to detect a nonclassical correlation of this state.
}\end{ob}
The use of a linear EnCE map for detecting nonclassical correlation
is intuitive and technically easy as we have seen.
There is, however, a case where the limitation of the linear EnCE maps
is clearly observed as described below.
\begin{rem}{\rm
There are states called {\it one-way classically correlated}
(1WCC) states \cite{H05}, in the form
\[
\rho_{\rm 1WCC}=
\sum_i|i^{\rm x}\rangle\langle i^{\rm x}|\otimes \sigma_i^{\rm y}
\]
with $|i^{\rm x}\rangle$ a CONS of ${\rm x}={\rm A~or~B}$ and ${\rm y}$
the remaining system; $\sigma_i^{\rm y}$ (unnormalized) density
operators acting on ${\rm y}$, dependent on the index $i$.
Such a state may have no product eigenbasis but testing a change in
the eigenvalues under
$I^{\rm x}\otimes\Lambda_{\rm EnCE}^{\rm y}$ for a single side
is not enough to detect it. Therefore we need to test for both
$\mathrm{(x,y)=(A,B)~and~(x,y)=(B,A)}$.
}\end{rem}
\begin{pro}
One cannot detect a nonclassical correlation of a one-way classically
correlated (1WCC) state using a linear EnCE map.
\end{pro}
\proof{
It is easy to find that
applying a partial transposition to $\rho_{\rm 1WCC}$
results in either $(U_*^{\rm x}\otimes I^{\rm y})~\rho_{\rm 1WCC}~
(U_*^{\rm x}\otimes I^{\rm y})^\dagger$ or
$(U_*^{\rm x}\otimes I^{\rm y})^\dagger~\rho_{\rm 1WCC}^*~
(U_*^{\rm x}\otimes I^{\rm y})$
with $U_*^{\rm x}=\sum_i (|i^{\rm x}\rangle^*)\langle i^{\rm x}|$. 
In addition, any partial unitary transformation preserves the
eigenvalues of $\rho_{\rm 1WCC}$.
By Proposition\ \ref{pro1}, the proof is completed.
}\\%end of proof
This proposition suggests that we need to search for nonlinear
EnCE maps for a wider range of detection than that of linear ones.
The definition involving both linear and nonlinear ones should be newly
given in consistent with Proposition\ \ref{pro2}.
\begin{de}
An EnCE map $\Lambda_{\rm EnCE}$ (that can be nonlinear)
should have the following properties.
\begin{itemize}
\item[(i)]
For any density matrix $\rho=\sum_{k=1}^d e_k|v_k\rangle\langle v_k|$,\\
$\Lambda_{\rm EnCE}:\rho\mapsto
\sum_{k=1}^d e_k|{v_k}'\rangle\langle {v_k}'|$ where
$\{|v_k\rangle\}$ and $\{|{v_k}'\rangle\}$ are both CONSs; $e_k$ are the
eigenvalues.\footnote{This is a property of any (possibly nonlinear) EP map.}
\item[(ii)]
For any bipartite density matrix $\rho^{\rm AB}_{\rm PE}$
with a product eigenbasis, written as (\ref{eq_rho_prod}),\\
$I^{\rm A}\otimes\Lambda_{\rm EnCE}^{\rm B}:\rho^{\rm AB}_{\rm PE}\mapsto
\widehat{\rho^{\rm AB}}$ where $\widehat{\rho^{\rm AB}}$ is an Hermitian
matrix with the set of the eigenvalues same as that of
$\rho^{\rm AB}_{\rm PE}$.
\item[(iii)]
For some bipartite density matrix $\sigma^{\rm AB}$
having no product eigenbasis,
$I^{\rm A}\otimes \Lambda_{\rm EnCE}^{\rm B}$ maps
it to an Hermitian matrix with the set of the eigenvalues
different from that of $\sigma^{\rm AB}$.
\end{itemize}
\end{de}
%Although we use this definition in this contribution, one may use a
%one-to-many relation in addition to the usual map relations, namely,
%many-to-one and one-to-one relations between the input state and
%the resultant matrix.
%Here we consider a usual map and
%do not employ a one-to-many relation.

We find that there is, in fact, a useful nonlinear EnCE map.
To define it, we first introduce the specially-designed nonlinear map
$\Gamma_x$.
\begin{de}
A nonlinear map $\Gamma_x$ acting on a (possibly unnormalized) quantum
state $\rho$ is defined as follows.
\[
\begin{array}{l}
\Gamma_x: \rho\mapsto \sqrt{(\rho\rho^{x-1})({\rm h.c.})}=\rho^x,\\
I^{\rm A}\otimes\Gamma_x^{\rm B}:
\rho^{\rm AB}\mapsto
\sqrt{
\{\rho^{\rm AB}[I^{\rm A}\otimes ({\rm Tr_A}\rho^{\rm AB})^{x-1}]\}
\{{\rm h.c.}\}},
\end{array}
\]
where $x\in{\bf R}$; the square root is positive and
${\rm h.c.}$ stands for Hermitian conjugate (conjugate transpose).\\
{}[Here, the inverse of a density matrix $\rho=\sum_k
e_k|v_k\rangle\langle v_k|$, with $(e_k, |v_k\rangle)$ the pair of an
eigenvalue and the (normalized) corresponding eigenvector, is defined as
$\rho^{-1}\equiv\sum_{k,~c_k\not = 0}c_k^{-1}|v_k\rangle\langle v_k|$.]
\end{de}
This is an extension of the $x$th power of a matrix.
Note that $(I^{\rm A}\otimes\Gamma_x^{\rm B})\rho^{\rm AB}$ 
is a quantum state (positive Hermitian matrix) because,
for positive Hermitian matrices $A$ and $B$, $(AB)({\rm h.c.})=ABBA$
is a positive Hermitian matrix.

A nonlinear EnCE map is now defined by using $\Gamma_x$.
\begin{de}\label{defP}
A nonlinear EnCE map $\mathcal{P}_x$ acting on a quantum state $\rho$ is
 defined as follows.
\[
\begin{array}{l}
\mathcal{P}_x: \rho^{\rm AB}\mapsto \Gamma_{1/x}\Gamma_{x}\rho^{\rm AB}=\rho^{\rm AB},\\
I^{\rm A}\otimes\mathcal{P}_x^{\rm B}: \rho^{\rm AB}\mapsto
(I^{\rm A}\otimes\Gamma_{1/x}^{\rm B})(I^{\rm A}\otimes\Gamma_{x}^{\rm B})\rho^{\rm AB},
\end{array}
\]
where $x\in{\bf R}, x\not = 1$.
\end{de}
Of course, $(I^{\rm A}\otimes\mathcal{P}_x^{\rm B})\rho^{\rm AB}$ 
is a quantum state (the trace is not preserved in general).

The map $\mathcal{P}_x$ is useful for detecting nonclassical
correlation because we have the following theorem.
\begin{thm}
The equations
$(I^{\rm A}\otimes\mathcal{P}_x^{\rm B})\rho^{\rm AB}
=(\mathcal{P}_x^{\rm A}\otimes I^{\rm B})\rho^{\rm AB}
=\rho^{\rm AB}$ hold if $\rho^{\rm AB}$ has a product eigenbasis.
\end{thm}
\proof{
For a bipartite state with a product eigenbasis (PE),
$\rho^{\rm AB}_{\rm PE}=\sum_{ij}c_{ij}|u_i\rangle^{\rm A}\langle u_i|\otimes |v_j\rangle^{\rm B}\langle v_j|$,
we have ${\rm Tr_A}\rho^{\rm AB}_{\rm PE}=\sum_l(\sum_kc_{kl})|v_l\rangle^{\rm B}\langle v_l|$.
Thus
\[
\begin{array}{rl}
\widetilde{\rho^{\rm AB}}&\equiv(I^{\rm A}\otimes \Gamma_x^{\rm B})\rho^{\rm AB}_{\rm PE}
=\sqrt{\{\rho^{\rm AB}_{\rm PE}[I^{\rm A}\otimes({\rm Tr_A}\rho^{\rm AB}_{\rm PE})^{x-1}]\}\{{\rm h.c.}\}}\\
&=\sum_{j,~f(j)\not = 0}f(j)^{x-1}\sum_ic_{ij}|u_i\rangle^{\rm A}\langle u_i|\otimes |v_j\rangle^{\rm B}\langle v_j|
\end{array}
\]
with $f(j)=\sum_kc_{kj}$.
For this matrix, we have
\[
{\rm Tr_A}\widetilde{\rho^{\rm AB}}=\sum_{t,~f(t)\not = 0}
f(t)^{x-1}f(t)|u_t\rangle^{\rm B}\langle u_t|
=\sum_{t,~f(t)\not = 0}
f(t)^x|u_t\rangle^{\rm B}\langle u_t|.
\]
Thus
\[
\begin{array}{rl}
\widetilde{\widetilde{\rho^{\rm AB}}}&\equiv(I^{\rm A}\otimes \Gamma_{1/x}^{\rm B})\widetilde{\rho^{\rm AB}}
=\sqrt{\{\widetilde{\rho^{\rm AB}}[I^{\rm A}\otimes({\rm Tr_A}\widetilde{\rho^{\rm AB}})^{(1-x)/x}]\}\{{\rm h.c.}\}}\\
&=\sum_{j,~f(j)\not = 0}f(j)^{x-1}f(j)^{1-x}\sum_ic_{ij}|u_i\rangle^{\rm A}\langle u_i|\otimes |v_j\rangle^{\rm B}\langle v_j|\\
&=\rho^{\rm AB}.
\end{array}
\]
This proves that $(I^{\rm A}\otimes\mathcal{P}_x^{\rm B})\rho^{\rm AB}_{\rm PE}=\rho^{\rm AB}_{\rm PE}$.
It is easy to show that $(\mathcal{P}_x^{\rm A}\otimes I^{\rm B})\rho^{\rm AB}_{\rm PE}=\rho^{\rm AB}_{\rm PE}$
in the same way.
}\\%end of proof
Here is a simple example to use this map for detecting nonclassical
correlation. For the bipartite state $\rho_{0+}$ which has been
introduced in (\ref{eq0+}),
$(I\otimes\mathcal{P}_{2})\rho_{0+}$ has the eigenvalues (approximately)
$0.826$, $0.375$, and (strictly) $0$ (with multiplicity two) which
are different from the eigenvalues of $\rho_{0+}$, $3/4$, $1/4$, and $0$
(with multiplicity two), except $0$'s. Therefore $\rho_{0+}$ has
no product eigenbasis.

\section{Quantification of nonclassical correlation}\label{sec4}
For the next step, we define a measure of nonclassical correlation
based on the theory of EnCE maps we have seen.
Note that, according to the definition, the set of the classically
correlated states is a nonconvex subset of the set of the separable states.
Thus it is not motivating to impose convexity on a measure of
nonclassical correlation. We may, however, impose a family of additivity
properties \cite{B03}. In particular, subadditivity is assessed in the
following. We begin with non-subadditive measures and later introduce
subadditive measures. A strategy to extend the detection ability of
subadditive measures is described.

\subsection{Non-subadditive measures}\label{sec4-1}
We first define a non-subadditive measure of nonclassical correlation
for a given EnCE map $\Lambda_{\rm EnCE}$ as follows.
Suppose we want to quantify a nonclassical correlation of a bipartite
system AB described by a density matrix $\rho^{\rm AB}$.
Then, we may consider the quantity with
subscript R (L) indicating that the right (left) component is acted by
$\Lambda_{\rm EnCE}$:
\[
 D_{\rm R,L}(\Lambda_{\rm EnCE},\rho^{\rm AB})=\sum_{s}|e_s-{e'}_s|,
\]
where $e_s$'s are the eigenvalues of $\rho^{\rm AB}$
while ${e'}_s$'s are those of
$(I^{\rm A}\otimes\Lambda_{\rm EnCE}^{\rm B})\rho^{\rm AB}$ for ``{\rm R}''
[$(\Lambda_{\rm EnCE}^{\rm A}\otimes I^{\rm B})\rho^{\rm AB}$
for ``${\rm L}$''];
$e_s$'s and ${e'}_s$'s are aligned, say, in the descending order.
We may use the transposition $\Lambda_{\rm T}$
or the map $\mathcal{P}_x$ defined in Definition \ref{defP} for
$\Lambda_{\rm EnCE}$. It is obvious that $D$ vanishes if
$\rho^{\rm AB}$ has a product eigenbasis.

We can easily calculate
$D_{\rm R,L}(\Lambda_{\rm EnCE},\rho^{\rm AB})$.
For example, it is easy to calculate\footnote{
The calculation is as follows.
$D_{\rm R}(\Lambda_{\rm T},\rho_{\rm ps})
= [(1+3p)/4 - (1+p)/4]
+ 2 \times [(1+p)/4 
-(1-p)/4] + [(1-p)/4-(1-3p)/4] 
= 2p$.
}~
$D_{\rm R}(\Lambda_{\rm T},\rho_{\rm ps})=2p$ for the two-qubit
state $\rho_{\rm ps}$ defined in (\ref{eqps}).

Another simple example is the quantification of a nonclassical
correlation for the bipartite state $\rho_{0+}$ which has been
introduced in (\ref{eq0+}).
The eigenvalues of $\rho_{0+}$ are $3/4$, $1/4$, and $0$
(with multiplicity two).
The quantity $D_{\rm R}(\Lambda_{\rm T},\rho_{0+})$ vanishes 
because $(I\otimes \Lambda_{\rm T})\rho_{0+}=\rho_{0+}$.
In contrast, $D_{\rm R}(\mathcal{P}_{2},\rho_{0+})$ does not vanish:
as we have computed in an example in the previous section,
$(I\otimes\mathcal{P}_{2})\rho_{0+}$ has the eigenvalues (approximately)
$0.826$, $0.375$, and (strictly) $0$ (with multiplicity two). Thus 
$D_{\rm R}(\mathcal{P}_{2},\rho_{0+})\simeq 0.201$.

The last example is to clarify that $D_{\rm R,L}$ is not subadditive.
For $|\psi\rangle=(|00\rangle+|11\rangle)/\sqrt{2}$,
$(I^{\rm A}\otimes\Lambda_{\rm T}^{\rm B})|\psi\rangle^{\rm
AB}\langle\psi|$ has the nondegenerate eigenvalue $-1/2$ and
the degenerate eigenvalue $1/2$ with multiplicity three.
The eigenvalues of $|\psi\rangle\langle\psi|$ are $1$ and $0$
with multiplicity three. Thus
$D_{\rm R}(\Lambda_{\rm T},|\psi\rangle\langle\psi|)=2$.
It is now easy to find the eigenvalues of 
$(I^{\rm AC}\otimes\Lambda_{\rm T}^{\rm BD})|\psi\rangle^{\rm
AB}\langle\psi|\otimes |\psi\rangle^{\rm CD}\langle\psi|$, which
are $-1/4$ with multiplicity six and $1/4$ with multiplicity ten.
This results in $D_{\rm R}(\Lambda_{\rm T},|\psi\rangle\langle\psi|
\otimes |\psi\rangle\langle\psi|)=9/2$. This value is larger than
$2\times D_{\rm R}(\Lambda_{\rm T},|\psi\rangle\langle\psi|)=4$,
indicating that subadditivity does not hold.

\subsection{Subadditive measures}\label{sec4-2}
It has been shown that the measures $D_{\rm R,L}$ introduced above
are neither additive nor subadditive.
Additive or subadditive measures are desirable if one needs to compare
systems with different dimensions. Here, subadditive measures are
introduced. Let us formally begin with the definition of subadditivity \cite{B03}.
\begin{de}
Let $F(\rho^{\rm AB})_{\rm A|B}$ be a measure of correlation between
subsystems A and B of a bipartite system AB, where ${\rm A|B}$ denotes
splitting between A and B. Then, $F(\rho^{\rm AB})_{\rm A|B}$ is called
a subadditive measure if and only if the relation
$F(\rho^{\rm AB}\otimes\sigma^{\rm CD})_{\rm AC|BD}
\le F(\rho^{\rm AB})_{\rm A|B} + F(\sigma^{\rm CD})_{\rm C|D}
$ holds for density matrices $\rho^{\rm AB}$ and $\sigma^{\rm CD}$
of systems ${\rm AB}$ and ${\rm CD}$ in general.
\end{de}

We find that the following quantities $Q_{\rm R}$ and $Q_{\rm L}$
satisfy the subadditivity condition if we choose the map
$\Lambda_{\rm EnCE}$ properly.
We define them as
\begin{equation}\label{eqQ}
 Q_{{\rm R},{\rm L}}(\Lambda_{\rm EnCE},\rho^{\rm AB})=
-\log_2\left(\frac{1}{N}\sum_s\sqrt{e_s\widetilde{e_s}}\right),
\end{equation}
where $e_s$'s are the eigenvalues of $\rho^{\rm AB}$ and
$\widetilde{e_s}$'s are the absolute values of the eigenvalues of
$(I^{\rm A}\otimes \Lambda_{\rm EnCE}^{\rm B})\rho^{\rm AB}$
for ``${\rm R}$'' [
$(\Lambda_{\rm EnCE}^{\rm A}\otimes I^{\rm B})\rho^{\rm AB}$
for ``${\rm L}$'']; $e_s$'s and
$\widetilde{e_s}$'s are both sorted, say, in descending order;
$N=\sqrt{\sum_{s}\widetilde{e_s}}$ is a normalization factor
which guarantees $Q_{{\rm R},{\rm L}}\ge 0$.
The measures $Q_{{\rm R},{\rm L}}$ vanish if
$\{e_s\}=\{\widetilde{e_s}/N^2\}$.
As for subadditivity, we can prove the following proposition.
\begin{pro}\label{prosub}
The measure $Q_{{\rm R}}(\Lambda_{\rm EnCE},\rho^{\rm AB})$
is subadditive if the set of the absolute values of the eigenvalues of
$(I^{\rm AC}\otimes\Lambda_{\rm EnCE}^{\rm BD})
(\rho^{\rm AB}\otimes\sigma^{\rm CD})$
is given by $\{\widetilde{a_j}\widetilde{b_k}\}_{jk}$ where
$\widetilde{a_j}$ and $\widetilde{b_k}$ are the absolute values
of the eigenvalues of
$(I^{\rm A}\otimes\Lambda_{\rm EnCE}^{\rm B})\rho^{\rm AB}$
and those of
$(I^{\rm C}\otimes\Lambda_{\rm EnCE}^{\rm D})\sigma^{\rm CD}$,
respectively.
\end{pro}
\proof{
The proof consists of two steps (i) and (ii).\\
(i) Consider the two sequences $\{p_i\}_{i=1}^{d}$ and
$\{q_i\}_{i=1}^{d}$ of nonnegative real numbers $p_i$ and $q_i$.
Suppose they are sorted: $p_1\ge p_2\ge \cdots \ge p_d$
and $q_1\ge q_2\ge \cdots \ge q_d$. Then,
the fidelity $\sum_{i=1}^d \sqrt{p_iq_i}$ for these sorted sequences
is larger than that for any two unsorted sequences whose entries are
 $p_i$'s and $q_i$'s, respectively. This is because,
for real numbers $a_1$, $a_2$, $b_1$, and $b_2$ such that $a_1\ge a_2$ and 
$b_1\ge b_2$, the relation $a_1b_1+a_2b_2\ge a_1b_2+a_2b_1$ holds.\\
(ii) Let us write the eigenvalues of $\rho^{\rm AB}$ as $a_j$ and
those of $\sigma^{\rm CD}$ as $b_k$.
The fidelity $F'=\sum_{j=1}^{d^{\rm A}}\sum_{k=1}^{d^{\rm B}}
\sqrt{(a_jb_k)(\widetilde{a_j}\widetilde{b_k})}/N$ with
$N=\sqrt{\sum_{jk}\widetilde{a_j}\widetilde{b_k}}$ involves two
possibly unsorted sequences $\{(a_jb_k)\}$ and
$\{(\widetilde{a_j}\widetilde{b_k})\}$; these are unsorted in general
even when $\{a_j\}$, $\{b_k\}$, $\{\widetilde{a_j}\}$, and
$\{\widetilde{b_k}\}$ are individually sorted.
Let us write the fidelity after sorting $\{(a_jb_k)\}$ and
$\{(\widetilde{a_j}\widetilde{b_k})\}$ as $F$. Then,
$F'\le F\le 1$ holds according to the fact (i).
Therefore, $0\le -\log_2 F\le -\log_2 F' = -\log_2 F_a -\log_2 F_b$ holds
with $F_a = \sum_j \sqrt{a_j\widetilde{a_j}}/\sqrt{\sum_j\widetilde{a_j}}$
and $F_b = \sum_k \sqrt{b_k\widetilde{b_k}}/\sqrt{\sum_k\widetilde{b_k}}$,
where $\{a_j\}$, $\{b_k\}$, $\{\widetilde{a_j}\}$, and
$\{\widetilde{b_k}\}$ are individually sorted.
}\\%end of proof
It is trivial to find a similar condition for 
$Q_{{\rm L}}(\Lambda_{\rm EnCE},\rho^{\rm AB})$ to be subadditive.
In addition, it is clear that $Q_{\rm R,L}$ vanish if
$\rho^{\rm AB}$ has a product eigenbasis. These measures are a sort
of logarithmic fidelity and are reminiscent of logarithmic
negativity \cite{neg:i,neg:ii,neg:iii}.
We find that choosing $\Lambda_{\rm EnCE}$ from the maps
$\Lambda_{\rm T}$ and $\mathcal{P}_{x}$ introduced in the previous
section satisfies the condition of Proposition\ \ref{prosub} as
we prove below.
As for other additivity properties, $Q_{\rm R,L}$ is not additive or
weakly additive in general owing to sortings of the eigenvalues.
This is clear from the following example: For the state $\rho_{\rm ps}$
defined in (\ref{eqps}), with $p$ set to $1/3$, we have
$Q_{\rm R}(\Lambda_{\rm T},
\rho_{{\rm ps},p=1/3}^{\rm AB}\otimes
\rho_{{\rm ps},p=1/3}^{\rm CD})_{\rm AC|BD}
=-\log_2 (5/18+\sqrt{3}/3)\simeq 0.226$; this is less than
$2\times Q_{\rm R}(\Lambda_{\rm T},\rho_{{\rm ps},p=1/3}^{\rm AB})_{\rm A|B}
=-2\log_2(\sqrt{6}/6+\sqrt{2}/3)\simeq 0.370$.
%
% For $\rho_{0+}$,
% \rho^{\rm AB} has eigs 3/4,1/4,0,0
% I\otimes\mathcal{P}_x \rho^{\rm AB} has eigs 0.8256,0.3753,0,0
% 
% (\rho^{\rm AB}\otimes\rho^{\rm CD}) has eigs 9/16,3/16,3/16,1/16,0,...,0
% I^{\rm AC}\otimes\mathcal{P}_x^{\rm BD}(...) has eigs 0.6816,0.3099,
% 0.3099,0.1409,0,...,0
%
% Thus Q_{R}(\mathcal{P}_x,
%\rho_{0+}^{\rm AB}\otimes\rho_{0+}^{\rm CD})_{\rm AC|BD}
% = -log_2((1/sqrt(1.4423)) x (0.6192+0.2411+0.2411+0.0938)) = 0.006934
%
%   Q_{R}(\mathcal{P}_x,\rho_{0+}^{\rm AB})_{\rm A|B}
% = -log_2((1/sqrt(1.2009247)) x (0.78689044 + 0.3063206)) = 0.0035009129
%
% 2 x Q_{R}(\mathcal{P}_x,\rho_{0+}^{\rm AB})_{\rm A|B} = 0.0070018
%
% So, Q_R for AC|BD < Q_R for A|B + Q_R for C|D for \rho_{0+}^{\otimes 2}.
%

As we have mentioned above, one map that makes the measure $Q_{\rm R,L}$
subadditive is the transposition $\Lambda_{\rm T}$. The subadditivity is
easily verified according to the fact that
$I^{\rm AC}\otimes\Lambda_{\rm T}^{\rm BD}
=(I^{\rm A}\otimes\Lambda_{\rm T}^{\rm B})(I^{\rm C}\otimes\Lambda_{\rm
T}^{\rm D})$. The condition on the set of eigenvalues stated in
Proposition \ref{prosub} is obviously satisfied.
In addition, the measures $Q_{\rm R,L}(\Lambda_{\rm T},
\rho^{\rm AB})$ are invariant under local unitary operations.
Its invariance under local unitary operations (say,
$U^{\rm B}$) follows from
$(I^{\rm A} \otimes \Lambda_{\rm T}^{\rm B})
(I^{\rm A} \otimes U^{\rm B} \rho^{\rm AB} I^{\rm A} \otimes {U^\dagger}^{\rm B})
= (I^{\rm A} \otimes {U^*}^{\rm B}) (I^{\rm A} \otimes
\Lambda_{\rm T}^{\rm B} \rho^{\rm AB}) (I^{\rm A} \otimes
{{U^*}^\dagger}^{\rm B}).$

It is also easy to find that $\mathcal{P}_{x}$, the map defined in Definition
\ref{defP}, makes $Q_{\rm R,L}$ subadditive.
Because ${\rm Tr_{AC}}\rho^{\rm AB}\otimes\rho^{\rm CD}=
({\rm Tr_{A}}\rho^{\rm AB})\otimes ({\rm Tr_{C}}\rho^{\rm CD})$,
we have
\[\begin{array}{rl}
&(I^{\rm AC}\otimes\Gamma_x^{\rm BD})(\rho^{\rm AB}\otimes\sigma^{\rm CD})\\
&=\sqrt{\{\rho^{\rm AB} [I^{\rm A}\otimes ({\rm Tr_A}\rho^{\rm AB})^{x-1}]
\otimes \sigma^{\rm CD} [I^{\rm C}\otimes ({\rm Tr_C}\sigma^{\rm CD})^{x-1}]\}
\{{\rm h.c.}\}}\\
&=(I^{\rm A}\otimes \Gamma_x^{\rm B})\rho^{\rm AB}
\otimes (I^{\rm C}\otimes \Gamma_x^{\rm D})\sigma^{\rm CD}.
\end{array}
\]
Hence, $(I^{\rm AC}\otimes \mathcal{P}^{\rm BD})(\rho^{\rm
AB}\otimes\sigma^{\rm CD})=(I^{\rm A}\otimes\mathcal{P}_x^{\rm
B})\rho^{\rm AB}\otimes (I^{\rm C}\otimes\mathcal{P}_x^{\rm
D})\sigma^{\rm CD}$ holds.
Therefore, by Proposition \ref{prosub}, we find that
$Q_{{\rm R}}(\mathcal{P}_{x},\rho^{\rm AB})$ is subadditive [and
$Q_{{\rm L}}(\mathcal{P}_{x},\rho^{\rm AB})$ either]. In addition,
it is invariant under local unitary operations because
$I^{\rm A,B}\otimes\Gamma_{x}^{\rm B,A}$ commutes with a
local unitary transformation, by its definition.

The problem is that $Q_{\rm R}$ and $Q_{{\rm L}}$ are different in
general. To solve this problem, we suggest using the average
\begin{equation}\label{eqQtil}
\widetilde{Q}(\Lambda_{\rm EnCE},\rho^{\rm AB})=
\frac{
Q_{{\rm R}}(\Lambda_{\rm EnCE},\rho^{\rm AB})
+Q_{{\rm L}}(\Lambda_{\rm EnCE},\rho^{\rm AB})
}{2}.
\end{equation}
This becomes subadditive and invariant under local unitary operations
if both $Q_{\rm R}$ and $Q_{{\rm L}}$ are subadditive
and invariant under local unitary operations. It is easy to find
that $\Lambda_{\rm T}$ and $\mathcal{P}_{x}$ are both useful for
this purpose.

\subsection{Computational complexity}\label{sec4-3}
An advantage of using the measure $Q_{\rm R,L}$ defined in (\ref{eqQ})
and the measure $\tilde Q$ defined in (\ref{eqQtil}) is their
relatively small computational cost when $\Lambda_{\rm T}$ or
$\mathcal{P}_x$ is chosen for $\Lambda_{\rm EnCE}$.

Consider a bipartite system with the dimension $d^{\rm A}$ ($d^{\rm B}$)
of the Hilbert space of its subsystem ${\rm A}$ (${\rm B}$).
For a density matrix $\rho^{\rm AB}$, 
$(I\otimes\Lambda_{\rm T})\rho^{\rm AB}$ is computed with
$O({d^{\rm A}}^2{d^{\rm B}}^2)$ basic floating-point operations.
This is less than the cost of diagonalization of $(I\otimes\Lambda_{\rm
T})\rho^{\rm AB}$, which takes $O({d^{\rm A}}^3{d^{\rm B}}^3)$ basic
floating-point operations.
The computation of $(I\otimes\mathcal{P}_{x})\rho^{\rm AB}$ is
a little expensive because it involves a square root of a matrix.
The cost of computing $(I\otimes\mathcal{P}_{x})\rho^{\rm AB}$ is
$O({d^{\rm A}}^3{d^{\rm B}}^3)$, same as the cost of diagonalizing
$(I\otimes\mathcal{P}_{x})\rho^{\rm AB}$.

Once the eigenvalues of $(I\otimes\Lambda_{\rm EnCE})\rho^{\rm AB}$
is computed, it takes only $O(d^{\rm A}d^{\rm B})$ basic floating-point 
operations to compute  $Q_{\rm R,L}$ and $\tilde Q$.
Therefore, the time complexity of computing these measures is
$O({d^{\rm A}}^3{d^{\rm B}}^3)$ when $\Lambda_{\rm T}$ or
$\mathcal{P}_x$ is chosen.

\subsection{Extending the detection range}\label{sec4-4}
One might be curious if $\mathcal{P}_x$, the map defined in Definition
\ref{defP}, is more useful than the transposition $\Lambda_{\rm T}$
in detecting nonclassical correlation by using the measure $\widetilde{Q}$
defined in (\ref{eqQtil}).
First, $\widetilde{Q}(\Lambda_{\rm T},\rho^{\rm AB}_{0+})$
vanishes while
$\widetilde{Q}(\mathcal{P}_2,\rho^{\rm AB}_{0+})\simeq 7.00\times
10^{-3}$ for the state $\rho_{\rm 0+}$ defined in (\ref{eq0+}).
Second, $\widetilde{Q}(\Lambda_{\rm T},|\psi\rangle^{\rm AB}\langle\psi|)
=1$ while $\widetilde{Q}(\mathcal{P}_x,|\psi\rangle^{\rm
AB}\langle\psi|)$ vanishes for
$|\psi\rangle=(|00\rangle+|11\rangle)/\sqrt{2}$.
Therefore, generally speaking,
$\widetilde{Q}(\mathcal{P}_x,\rho^{\rm AB})$ is neither stronger
nor weaker than $\widetilde{Q}(\Lambda_{\rm T},\rho^{\rm AB})$.
One may further claim that $\mathcal{P}_x$ is not very useful because
it vanishes for the Bell state. Nevertheless, this is not a serious
drawback as we have a quick solution as follows.

There is a way to utilize these measures to produce a stronger measure.
Suppose we have non-negative, subadditive, and local-unitary-invariant
measures $M_1, ..., M_N$. Then, the weighted average
$\sum_k w_k M_k$ with $w_k>0$ is also a measure which is non-negative,
subadditive, and invariant under local unitary operations. It detects
nonclassical correlation for the states for which any one of $M_1,...,M_N$
is nonvanishing.

Thus we easily produce the stronger measure
\[
w_{\rm T} \widetilde{Q}(\Lambda_{\rm T},\rho^{\rm AB})
+ \sum_k w_k \widetilde{Q}(\mathcal{P}_{x_k},\rho^{\rm AB}) 
\]
with $x_k \in {\bf R}$, $x_k \not = 1$, and $w_{\rm T},w_k>0$.
This measure does not vanish for $\rho_{0+}$ and
$|\psi\rangle\langle\psi|$.

%%%%
% Discussions and conclusion
%%%%
\section{Concluding remarks}\label{sec6}
We have seen several different usages of the EnCE map theory. We believe
that this theory works as a useful template to detect and quantify
nonclassical correlation based on the Oppenheim-Horodecki separation of
classical/nonclassical correlations.
The EnCE map theory has been constructed in analogy to the PnCP map
theory in the present paper. One important difference between these theories
is that the class of EnCE maps includes nonlinear EnCE maps. This is
because linear EnCE maps are very limited due to the fact that any
linear EP map can be decomposed into unitary operations and a transposition
(Proposition\ \ref{pro1}). Nonlinearity of a map is not a significant
problem as far as $I\otimes\Lambda_{\rm EnCE}$ and $\Lambda_{\rm
EnCE}\otimes I$ are defined appropriately for an EnCE map $\Lambda_{\rm EnCE}$
in the way that $I\otimes \Lambda_{\rm EnCE}$ and $\Lambda_{\rm EnCE}
\otimes I$ preserve the eigenvalues of any state that has a product
eigenbasis. If one intends to rule out nonlinearity, a possible extension of
the theory is to go beyond the tacit assumption of the Hermiticity-preserving
property of a map. This will be studied in the future.

On the basis of the EnCE map theory, we have defined two subadditive
measures, $\widetilde{Q}(\Lambda_{\rm T},\rho^{\rm AB})$
and $\widetilde{Q}(\mathcal{P}_x,\rho^{\rm AB})$.
These are neither stronger nor weaker to each other in the detection
range, and not so strong as the measure by Piani {\em et al.}
\cite{PHH08} that is perfect in the detection range albeit intractable
in computational cost. The advantage of our measures is the
complexity: they are calculated within polynomial time in the dimension
of the Hilbert space. We have shown a way to relax the drawback
of the detection range; their weighted average is stronger than
themselves and remains subadditive as shown in Section\ \ref{sec4-4}.
A certain optimization over the weights and the choices of $x$'s
will be investigated in future work.

One might be curious about an extension of the measures to
multipartite splitting. This is achieved by
taking a minimum, maximum, or average of a measure over all
possible bipartite splittings of the multipartite system.
In considering the possible combinations of subsystems for a
bipartite splitting, we should be careful about the fact that having
product eigenbases
for ${\rm A|BC}$ splitting and ${\rm AB|C}$ splitting does not imply
having a product eigenbasis for ${\rm A|B|C}$ splitting.
A typical example is the state
$(|000\rangle\langle000|+|1+1\rangle\langle1+1|)/2$ with
$|+\rangle=(|0\rangle+|1\rangle)/{\sqrt{2}}$.
This state does not have a product eigenbasis for ${\rm A|B|C}$
splitting while it has for ${\rm A|BC}$ and ${\rm AB|C}$ splittings.
A proper claim is that having product eigenbases for
the ${\rm A|BC}$, ${\rm AB|C}$, and ${\rm AC|B}$ splittings implies
having a product eigenbasis for ${\rm A|B|C}$ splitting.
More generally, an $m$-partite state $\rho^{1...m}$ has a 
fully product eigenbasis if and only if $\rho^{1...m}$ has
a product eigenbasis for every possible bipartite splitting
separating $\{1,...,m\}$ into two sets.
The proof is given in Appendix \ref{app1}.
%%%
In addition, as a different direction to study the measures for a
multipartite system, one may seek for a monogamy property, namely, a sort
of restriction to a subsystem in the amount of correlation with other
subsystems when it has a correlation with a particular subsystem
(See, e.g., Ref.~\cite{KW04} and references therein). It is an open
problem if a measure in the form of (\ref{eqQ}) fulfills a certain
monogamy property alone or together with a different measure of classical
or nonclassical correlation.

In summary, a comprehensive framework, called the EnCE map theory,
to detect and quantify nonclassical correlation of a bipartite system
has been proposed. The average
logarithmic fidelity $\widetilde{Q}(\Lambda_{\rm EnCE},\rho^{\rm AB})$
has been introduced as a subadditive measure for a properly-chosen
EnCE map $\Lambda_{\rm EnCE}$. It is computable within polynomial time
in the dimension of the Hilbert space. A simple way to extend the
detection range by a collection of measures has been developed.

\nonumsection{Acknowledgements}
\noindent
A.S. and M.N. are supported by the ``Open Research Center'' Project
for Private Universities: matching fund subsidy from MEXT. R.R. is
supported by the FIRST program of JSPS. A.S., R.R., and M.N. are supported
by Grants-in-Aid for Scientific Research from JSPS (Grant Nos. 21800077,
1907329, and 19540422, respectively).

\nonumsection{References}
\noindent
\bibliographystyle{unsrt-mod}
\bibliography{refs_nonclassical}

\appendix{~Theorem on multipartite product eigenbasis}\label{app1}
\begin{thm}
An $m$-partite state $\rho^{1...m}$ has a
fully product eigenbasis if and only if $\rho^{1...m}$ has
a product eigenbasis for every possible bipartite splitting
separating $\{1,...,m\}$ into two sets.
\end{thm}
\proof{
It is trivial that $\rho^{1...m}$ has a product eigenbasis for
every possible bipartite splitting if it has a fully product eigenbasis.

Now we prove the converse. By lemma \ref{lem1} introduced below,
the density matrix has a product eigenbasis for the
$1|2|34...m$ splitting and that for the $12|3|4...m$ splitting.
The latter fact implies that $\rho^{1...m}$'s eigenbasis is a product of
the eigenbasis of the reduced density matrix $\rho^{12}$, that of
$\rho^3$, and that of $\rho^{4...m}$. The former fact implies that
the reduced density matrix $\rho^{12}$ has a product eigenbasis.
Therefore, $\rho^{1...m}$ has a product eigenbasis for the
$1|2|3|4...m$ splitting.

Next, we use the fact that $\rho^{1...m}$ has a product eigenbasis
for the $123|4|56...m$ splitting by lemma \ref{lem1}.
Now it is found that $\rho^{1...m}$ has a product eigenbasis for
the $1|2|3|4|56...m$ splitting.

Using the same logic continuously, the converse is proved.
}%end of proof

\begin{lemm}\label{lem1}
A tripartite density matrix $\rho^{\rm ABC}$ has a tripartite
product eigenbasis if and only if it has a bipartite product eigenbasis
for each of all the bipartite splittings. 
\end{lemm}
\proof{
It is trivial that $\rho^{\rm ABC}$ has a product eigenbasis for any
bipartite splitting if it has a tripartite product eigenbasis.

Now we prove the converse.
Having a bipartite product eigenbasis for any bipartite splitting
implies that 
\[
\begin{array}{rl}
\rho^{\rm ABC}&=\sum_{ij}a_{ij}|r_i\rangle^{\rm A}\langle r_i|
\otimes |s_j\rangle^{\rm BC}\langle s_j|\\
&=\sum_{kl} b_{kl}|t_k\rangle^{\rm AB}\langle t_k|
\otimes |u_l\rangle^{\rm C}\langle u_l|\\
&=\sum_{mn} c_{mn}|v_m\rangle^{\rm AC}\langle v_m|
\otimes |w_n\rangle^{\rm B}\langle w_n|,
\end{array}
\]
where
$|r_i\rangle^{\rm A}$, $|s_j\rangle^{\rm BC}$,
$|t_k\rangle^{\rm AB}$, $|u_l\rangle^{\rm C}$,
$|v_m\rangle^{\rm AC}$, and $|w_n\rangle^{\rm B}$
are eigenvectors of the reduced density matrices
of the indicated subsystems; $a_{ij}$, $b_{kl}$, and $c_{mn}$
are eigenvalues of $\rho^{\rm ABC}$.

This leads to that\\
(i) An eigenbasis of $\rho^{\rm ABC}$ is a product of
an eigenbasis of ${\rm Tr_{BC}}\rho^{\rm ABC}$ and
that of ${\rm Tr_A}\rho^{\rm ABC}$.\\
(ii) Matrix ${\rm Tr_A}\rho^{\rm ABC}$ is represented as
\[
\begin{array}{rl}
{\rm Tr_A}\rho^{\rm ABC}&=
\sum_{kl}b_{kl}\sigma_k^{\rm B}\otimes |u_l\rangle^{\rm C}\langle u_l|\\
&=\sum_{mn} c_{mn} |w_n\rangle^{\rm B}\langle w_n|
\otimes {\sigma'}_m^{\rm C}
\end{array}
\]
with $\sigma_k^{\rm B}={\rm Tr_A}|t_k\rangle^{\rm AB}\langle t_k|$
and ${\sigma'}_m^{\rm C}={\rm Tr_A}|v_m\rangle^{\rm AC}\langle v_m|$.\\
From (ii), we find that
\[\begin{array}{rl}
{\rm Tr_A}\rho^{\rm ABC}|w_x\rangle^{\rm B}|u_y\rangle^{\rm C}
&=(\sum_k b_{ky}\sigma_k^{\rm B}|w_x\rangle^{\rm B})|u_y\rangle^{\rm C}\\
&=|w_x\rangle^{\rm B}(\sum_m c_{mx}{\sigma'}_m^{\rm C}|u_y\rangle^{\rm C}).
\end{array}
\]
This implies that $(\sum_k b_{ky}\sigma_k^{\rm B}|w_x\rangle^{\rm B})
=p_{xy}|w_x\rangle^{\rm B}$ with $p_{xy}={}^{\rm C}\langle u_y|
\sum_m c_{mx}{\sigma'}_m^{\rm C}|u_y\rangle^{\rm C}$.
Hence $\{|w_x\rangle^{\rm B}|u_y\rangle^{\rm C}\}$ is an
eigenbasis of ${\rm Tr_A}\rho^{\rm ABC}$.
This fact and (i) complete the proof.
}%end of proof

\end{document}